\documentclass[12pt]{article}
\usepackage{amssymb,amsmath,bm}
\usepackage{epsf}
\usepackage{epsfig}
\usepackage{psfrag}
\usepackage{rotating}
\newcommand{\beq}{\begin{eqnarray}}
\newcommand{\eeq}{\end{eqnarray}}

\newcommand{\vcb}{|V_{cb}|}
\newcommand{\vtd}{|V_{td}|}

\newcommand{\vts}{|V_{ts}|}
\newcommand{\vus}{|V_{us}|}

\def\R1{\varepsilon_1}
\def\E8{\varepsilon_8}

\newcommand{\nn}{\nonumber}

\newcommand{\mtb}{\overline{m}_{\rm t}}
\newcommand{\mcb}{\overline{m}_{\rm c}}

\newcommand{\mw}{M_W}

\newcommand{\gev}{\, {\rm GeV}}
\newcommand{\mev}{\, {\rm MeV}}

\newcommand{\bea}{\begin{eqnarray}}
\newcommand{\eea}{\end{eqnarray}}
\newcommand{\bd}{\begin{displaymath}}
\newcommand{\ed}{\end{displaymath}}

\newcommand{\be}{\begin{equation}}
\newcommand{\ee}{\end{equation}}
\newcommand{\bi}{\begin{itemize}}
\newcommand{\ei}{\end{itemize}}
\newcommand{\ord}{{\cal O}}

\def\kpn{K^+\rightarrow\pi^+\nu\bar\nu}
\def\klpn{K_{\rm L}\rightarrow\pi^0\nu\bar\nu}

\renewcommand{\baselinestretch}{1.3}

\begin{document}
\thispagestyle{empty}
\phantom{xxx}
\vskip1truecm
\begin{flushright}
 TUM-HEP-626/06
\end{flushright}
\vskip1.5truecm

\begin{center}
 {\Large\bf Minimal Flavour Violation\vspace{-0.1cm}\\ 
Waiting for Precise\vspace{0.2cm}  Measurements\vspace{+0.1cm}\\
of \boldmath{$\Delta M_{s}$}, \boldmath{$S_{\psi\phi}$},
\boldmath{$A^s_\text{SL}$}, \boldmath{$|V_{ub}|$}, \boldmath{$\gamma$} and \boldmath{$B^0_{s,d}\to \mu^+\mu^-$} 
}
\end{center}

\vskip0.8truecm

\begin{center}
{\large\bf Monika Blanke, Andrzej J. Buras,\\ 
Diego Guadagnoli
and Cecilia Tarantino} 
\vspace{0.4truecm}

{\sl Physik Department, Technische Universit\"at M\"unchen,
D-85748 Garching, Germany}

\vspace{0.2truecm}

\end{center}

\centerline{\bf Abstract}
\noindent
We emphasize that the recent measurements of the $B^0_{s}-\bar B^0_{s}$ mass difference 
$\Delta M_{s}$ by the CDF and D{\O} collaborations 
offer an important {\it model
independent} test of  minimal flavour violation (MFV). The improved 
measurements of the angle $\gamma$ in the unitarity triangle and of 
$|V_{ub}|$ from tree level decays, 
combined with future accurate 
measurements of $\Delta M_{s}$, $S_{\psi K_S}$, $S_{\psi\phi}$, $Br(B_{d,s}\to\mu^+\mu^-)$, 
$Br(B\to X_{d,s}\nu\bar\nu)$, $Br(\kpn)$ and $Br(\klpn)$ and improved
values of the relevant non-perturbative parameters, will allow to test the
MFV hypothesis in a model independent manner to a high accuracy. 
In particular, the difference between 
the {\it reference} unitarity triangle obtained from tree level processes
 and the  {\it universal}
 unitarity triangle (UUT) in MFV models would signal either new flavour 
violating interactions and/or new local operators that are suppressed
in MFV models with low $\tan \beta$, with the former best tested
through $S_{\psi\phi}$ and $K_\text{L}\to\pi^0\nu\bar\nu$.  
 A brief discussion of non-MFV scenarios is
also given. In this context
we identify in the recent literature a relative sign error between Standard
Model and new physics contributions to $S_{\psi\phi}$, that has an impact on
the correlation between $S_{\psi\phi}$ and $A^s_\text{SL}$.
We point out that the ratios $S_{\psi \phi}/A^s_\text{SL}$ and $\Delta M_s/\Delta \Gamma_s$ will allow to determine $\Delta M_s/(\Delta M_s)^\text{SM}$. Similar proposals for the determination of $\Delta M_d/(\Delta M_d)^\text{SM}$ are also given.

\newpage

\section{Introduction}\setcounter{equation}{0}

The recent measurement of the $B^0_{s}-\bar B^0_{s}$ mass
difference by the CDF collaboration \cite{CDFnew} 
\be\label{CDF}
\Delta M_{s}=(17.33^{+0.42}_{-0.21}\pm 0.07)/{\rm ps}
\ee
and the two-sided bound by
the D{\O} collaboration \cite{Abazov:2006dm} $17/{\rm ps}\le \Delta M_{s}\le 21/{\rm ps}\; (90\%~~{\rm C.L.})$
provided still another constraint on the Standard Model (SM) and its 
extensions. In particular, the value of $\Delta M_{s}$ measured by the
CDF collaboration turned out to be surprisingly below the SM
predictions obtained from other constraints \cite{UTFIT,CKMFIT}
\be\label{DMsSM}
\left(\Delta
  M_s\right)^\text{SM}_\text{UTfit}=(21.5\pm2.6)/\text{ps},\qquad
\left(\Delta
  M_s\right)^\text{SM}_\text{CKMfitter}=\left(21.7^{+5.9}_{-4.2}\right)/\text{ps}.
\ee
The tension between \eqref{CDF} and \eqref{DMsSM} is
not yet significant, due to the sizable non-perturbative
uncertainties.
A consistent though slightly smaller value is found for the mass difference
directly from its SM expression \cite{BJW90}
\be
(\Delta M_s)^\text{SM}_\text{direct}=\dfrac{G_F^2}{6\pi^2}
\eta_B m_{B_s} \left(\hat B_{B_s} F_{B_s}^2\right) M_W^2 S(x_t) |V_{ts}|^2 = (17.8 \pm 4.8)/\text{ps}\,,
\label{DMsSMb}
\ee
with $|V_{ts}|=0.0409 \pm 0.0009$ and the other input parameters collected in Table \ref{tab:input}.

It should be
emphasized that the simplest extensions of the SM favoured  $\Delta
M_{s}>(\Delta M_s)^\text{SM}$. A notable exception is the MSSM with
minimal flavour violation (MFV) and large $\tan\beta$, where the
suppression of $\Delta M_s$ with respect to $(\Delta M_s)^\text{SM}$ has
been predicted \cite{BCRS}. In more complicated models, like the MSSM
with new flavour violating interactions \cite{Ciuchini:2006dx},
$\Delta M_s$ can be smaller or larger than $(\Delta M_s)^\text{SM}$.

In this paper we would like to emphasize that this new result
offers an important {\it model independent} test of 
models with MFV 
\cite{UUT,AMGIISST,Chivukula:1987py}, 
within the $B^0_d$ and
$B^0_s$ systems. We will summarize its implications for MFV 
models 
and discuss briefly non-MFV scenarios. The first version of our paper appeared few days before 
the announcement of the result in (\ref{CDF}) \cite{CDFnew}, which has
considerably reduced the uncertainties and prompted us to extend our analysis.

We will use first a constrained definition of MFV \cite{UUT}, 
to be called CMFV in what follows,
in which
\begin{itemize}
\setlength{\itemsep}{0pt plus1pt minus 1pt}
\setlength{\topsep}{0pt plus1pt minus 1pt}
\item
 flavour and CP violation is
 exclusively governed by the CKM matrix \cite{CKM}
\item
the structure of low energy operators is the same as in the SM.
\end{itemize}

The second condition introduces an additional constraint not 
present in the general formulation of \cite{AMGIISST}, 
but has the virtue that 
CMFV can be tested by means of
relations between various observables that are independent of the parameters 
specific to a given CMFV model \cite{UUT}. The violation of these relations 
would indicate the relevance of new low energy operators and/or the presence 
of new sources of flavour and CP violation, encountered for instance in general 
supersymmetric models \cite{GGMS}. The first studies of the implications 
of the $\Delta M_s$ experimental results on the parameters of such models can be found
in \cite{Ciuchini:2006dx,Endo:2006dm,Foster:2006ze,Cheung:2006tm,Baek:2006fq,He:2006bk} and the result in
\eqref{CDF} has been included in the analyses of the UTfit and
CKMfitter collaborations \cite{UTFIT,CKMFIT}.

Our paper is organized as follows: Section 2 is devoted entirely to
CMFV and $\Delta B=2$ transitions. In Section 3 we study the
implications of \eqref{CDF} on the CMFV relations between $\Delta B=1$
and  $\Delta B=2$ processes. In Section 4 we discuss briefly the tests
involving both $K$ and $B$ systems. In Section 5 we discuss the impact
of new operators still in the context of MFV. In Section 6 we
analyse some aspects of non-MFV scenarios, and in Section 7 we have a
closer look at the CP asymmetry $S_{\psi\phi}$ and its correlation
with $A^s_\text{SL}$. In
Section 8 we give a
brief summary of our findings.

\section{Basic Relations and their First Tests}\setcounter{equation}{0}

\begin{figure}[hbt]
\vspace{0.10in}
\centerline{
\epsfysize=1.8in
\epsffile{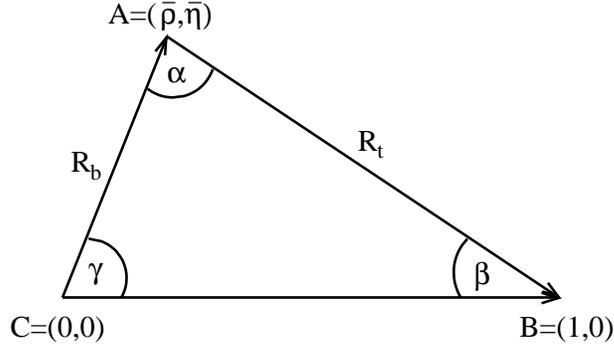}
}
\vspace{0.08in}
\caption{Unitarity Triangle.}\label{fig:utriangle}
\end{figure}

It will be useful to adopt the following sets of fundamental 
parameters related to the CKM matrix and the unitarity triangle 
shown in  Fig.~\ref{fig:utriangle}:
\begin{gather}\label{set1}
\vus\equiv\lambda,\qquad \vcb,\qquad R_b,\qquad \gamma,\\
\vus\equiv\lambda,\qquad \vcb,\qquad R_t,\qquad \beta.\label{set2}
\end{gather}

The following known expressions will turn out to be useful in what 
follows:
\begin{gather}
\label{2.94}
R_b \equiv \frac{| V_{ud}^{}V^*_{ub}|}{| V_{cd}^{}V^*_{cb}|}
= \sqrt{\bar\varrho^2 +\bar\eta^2}
= \left(1-\frac{\lambda^2}{2}\right)\frac{1}{\lambda}
\left| \frac{V_{ub}}{V_{cb}} \right|,\\
\label{2.95}
R_t \equiv \frac{| V_{td}^{}V^*_{tb}|}{| V_{cd}^{}V^*_{cb}|} =
 \sqrt{(1-\bar\varrho)^2 +\bar\eta^2}
=\frac{1}{\lambda} \left| \frac{V_{td}}{V_{cb}} \right|.
\end{gather}

While  set (\ref{set1}) can be determined entirely from tree level
decays and consequently independently of new physics contributions, the 
variables $R_t$ and $\beta$ in set (\ref{set2}) can only be determined in
one-loop induced processes and are therefore in principle sensitive to new
physics. It is the comparison between the values for the two sets of 
parameters determined in the respective processes, that offers a powerful 
test of CMFV, when the unitarity of the CKM matrix is imposed. One finds
then the relations
\be\label{VUBG}
R_b=\sqrt{1+R_t^2-2 R_t\cos\beta},\qquad
\cot\gamma=\frac{1-R_t\cos\beta}{R_t\sin\beta},
\ee
which are profound within CMFV for the following reasons. 
The quantities on the l.h.s. 
of (\ref{VUBG}) can be determined entirely in tree level processes, whereas 
the variables $\beta$ and $R_t$ from one-loop induced processes. The
important virtue of CMFV, to be contrasted with other extensions of the SM, 
is that the determination of 
$\beta$ and $R_t$ does not require the specification of a given CMFV model. 
In particular, determining $\beta$ and $R_t$ by means of
\be\label{R1}
\sin 2\beta= S_{\psi K_S}, 
\ee
\begin{eqnarray}\nonumber
R_t&=&\frac{\xi}{\lambda}\sqrt{\frac{\Delta M_d}{\Delta M_s}}
     \sqrt{\frac{m_{B_s}}{m_{B_d}}}
    \left[1-\lambda\xi\sqrt{\frac{\Delta M_d}{\Delta M_s}}
\sqrt{\frac{m_{B_s}}{m_{B_d}}}\cos\beta+\frac{\lambda^2}{2}+\ord(\lambda^4) 
\right]\\
\label{RRt}
&\approx& 0.923~\left[\frac{\xi}{1.23}\right] 
\sqrt{\frac{17.4/\text{ps}}{\Delta M_s}} 
\sqrt{\frac{\Delta M_d}{0.507/\text{ps}}}, 
\end{eqnarray}
where \cite{Hashimoto}
\be\label{xi}
\xi = 
\frac{\sqrt{\hat B_{B_s}}F_{B_s} }{ \sqrt{\hat B_{B_d}}F_{B_d}}=1.23\pm 0.06,
\ee
allows to construct the UUT \cite{UUT} for all CMFV models that
can be compared with the reference unitarity triangle \cite{refut} 
following from $R_b$ 
and $\gamma$. The difference between these two triangles signals new 
sources of flavour violation and/or new low energy operators 
 beyond the CMFV scenario. 
Here, $S_{\psi K_S}$ stands for the coefficient of 
$\sin(\Delta M_d t)$ in the mixing induced CP asymmetry in 
$B^0_d(\bar B^0_d)\to \psi K_S$ and, in obtaining the expression
(\ref{RRt}) for $R_t$, we have taken into account a small difference between $\vcb$ and $\vts$, 
that will play a role once the accuracy on $\xi$ and $\Delta M_s$ 
improves.

The values of the input parameters entering in \eqref{RRt} and used in
the rest of the paper are collected in Table \ref{tab:input}. In
particular, we take as lattice averages of $B$-parameters and
decay constants the values quoted in \cite{Hashimoto}, which combine
unquenched results obtained with different lattice actions.

\begin{table}[ht]
\renewcommand{\arraystretch}{1}\setlength{\arraycolsep}{1pt}
\center{\begin{tabular}{|l|l|}
\hline 
{\small $G_F=1.16637\cdot 10^{-5} \gev^{-2}$} & {\small$|V_{ub}|=0.00423(35)$} \\
{\small$\mw= 80.425(38)\gev$} &{\small $\vcb = 0.0416(7)$\hfill\cite{BBpage}}\\\cline{2-2}
{\small$\alpha=1/127.9$} &{\small$\lambda=0.225(1)$\hfill\cite{CKM05}} \\\cline{2-2}
{\small$\sin^2 \theta_W=0.23120(15)$} &{\small$F_{B_s} \sqrt{\hat B_{B_s}}= 262(35)\mev$}\\
{\small$m_\mu= 105.66\mev$} & {\small$\xi=1.23(6)$} \\
{\small$\Delta M_K= 3.483(6)\cdot 10^{-15}\gev$}&{\small $\hat B_{B_d}=1.28(10)$}\\
{\small$F_K= 159.8(15)\mev$} & {\small $\hat B_{B_s}=1.30(10)$}\\
{\small$m_{K^0}= 497.65(2)\mev$}\hfill\cite{PDG} & {\small $\hat B_{B_s}/\hat B_{B_d}=1.02(4)$\hfill\cite{Hashimoto}}\\\hline
{\small$m_{B_d}= 5.2793(7)\gev$} &{\small$\eta_1=1.32(32)$\hfill\cite{HNa}}\\ \cline{2-2}
{\small$m_{B_s}= 5.370(2)\gev$} & {\small$\eta_3=0.47(5)$\hfill\cite{HNb}}\\\cline{2-2}
{\small$\tau(B_d)= 1.530(9)\,\rm{ps}$} &{\small$\eta_2=0.57(1)$}\\
{\small$\tau(B_s)= 1.466(59)\,\rm{ps}$} &{\small$\eta_B=0.55(1)$\hfill\cite{BJW90}}\\\cline{2-2}
{\small$\tau(B_s)/\tau(B_d)=0.958(39)$}&{\small$\eta_Y=1.012(5)$\hfill\cite{BB99}}\\\cline{2-2}
{\small$\Delta M_d=0.507(5)/ \rm{ps}$} &{\small$\mcb= 1.30(5)\gev$}\\
{\small $S_{\psi K_S}=0.687(32)$ \hfill\cite{BBpage}}  &{\small$\mtb= 163.8(32)\gev$}\\
\hline
\end{tabular}  }
\caption {{Values of the experimental and theoretical
    quantities used as input parameters.}}
\label{tab:input} 
\renewcommand{\arraystretch}{1.0}
\end{table}

Until 
the recent measurement of $\Delta M_s$ in (\ref{CDF})  \cite{CDFnew}, 
none
of the relations in (\ref{VUBG})  could be tested in a 
model independent manner, even if the imposition of other constraints like 
$\varepsilon_K$ and separate information on $\Delta M_d$ and $\Delta M_s$ 
implied already interesting results for models with CMFV \cite{UTFIT,CKMFIT,MFVB}.
In particular in \cite{AMGIISST} the UUT has been constructed by using 
$\varepsilon_K$, $\Delta M_d$ and $\Delta M_s$ and treating the relevant 
one-loop function $S$ as a free parameter. A similar strategy has been used 
earlier in \cite{AJBRB} to derive a lower bound on $\sin 2\beta$ from CMFV. 
While such an approach is clearly legitimate, we think that using only 
quantities in which one has fully eliminated the dependence on new physics 
parameters allows a more transparent test of CMFV, and in the case of data 
indicating 
departures from CMFV, to identify clearly their origin.

With the measurement of $\Delta M_s$ 
in (\ref{CDF}) at hand, $S_{\psi K_S}$ and $\Delta M_d$ known 
very precisely \cite{BBpage}, we find using \eqref{R1} and \eqref{RRt}
\be\label{input}
(\sin 2 \beta)_\text{CMFV}= 0.687\pm 0.032, \qquad (R_t)_{\rm CMFV}=0.923\pm0.044,
\ee
and subsequently, using (\ref{VUBG}),
\be\label{RCMFV}
(R_b)_{\rm CMFV}=0.370\pm0.020, \qquad \gamma_{\rm CMFV}=(67.4\pm6.8)^\circ.
\ee
This should be compared with the values for
$R_b$ and $\gamma$ known from tree level semileptonic $B$ decays \cite{BBpage} and 
$B\to D^{(*)}K$ \cite{UTFIT}, respectively
\be\label{input2}
(R_b)_\text{true}=0.440\pm 0.037, \qquad \gamma_\text{true}=(71\pm16)^\circ.
\ee
The  relations 
in (\ref{VUBG}) can then be tested for the first time, even if the quality 
of the test is still not satisfactory. 
We have dropped in (\ref{input2})  the solution 
$\gamma=-(109\pm16)^\circ$ as it is inconsistent with $\beta >0$ within the 
MFV framework, unless the new physics contributions to the 
one-loop function $S$ in $B^0_d-\bar B^0_d$ mixing reverse its sign 
\cite{BF01}. Moreover, it is ruled out by the lower bound on $\Delta M_s$.

\begin{figure}\begin{minipage}{7cm}
\epsfig{file=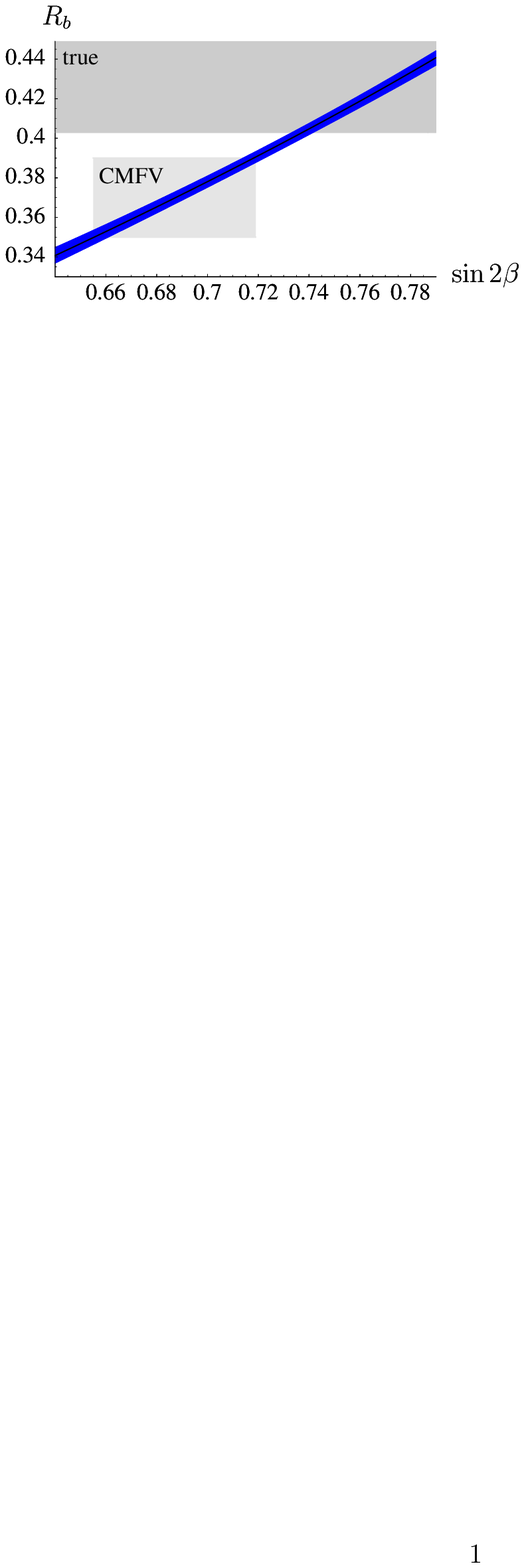,scale=1}\end{minipage}\hfill
\begin{minipage}{7.7cm}
\epsfig{file=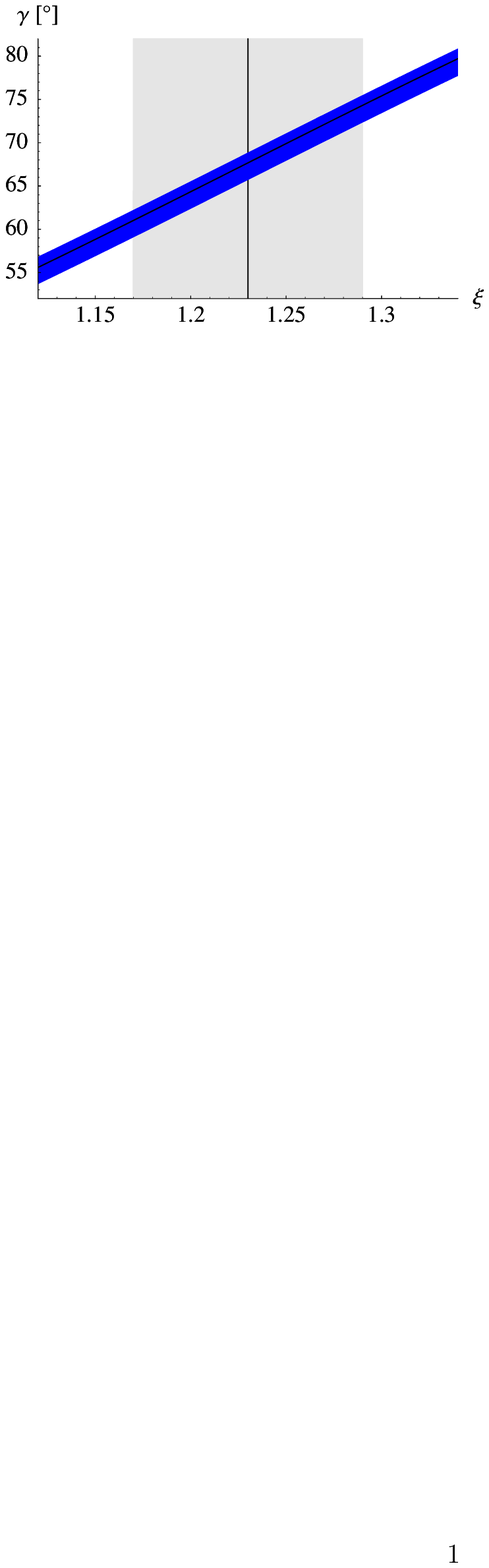,scale=1}\end{minipage}
\label{fig:Rb_and_gamma}\vspace{0.2cm}
\caption{$R_b$ and $\gamma$ in CMFV as functions  of $\sin 2\beta$ and
  $\xi$, respectively.}
\end{figure}

With future improved 
measurements of $\Delta M_s$, of $\gamma$ from $B\to D^{(*)}K$ and other 
tree level decays, a more accurate 
value for $R_b$ from $|V_{ub}/V_{cb}|$ and a more accurate value of $\xi$, 
the important tests of CMFV summarized in (\ref{VUBG}) 
will become effective.

In the left panel of Fig.~\ref{fig:Rb_and_gamma} we show 
$R_b$ as a function of 
$\sin 2\beta$  
for $\xi$ and $\Delta M_s$  varied in the ranges (\ref{xi}) and 
(\ref{CDF}) respectively. The lower part of the range (\ref{input2}) obtained for $R_b$ from
tree level semileptonic decays is also shown. This plot and the comparison of (\ref{RCMFV}) and 
(\ref{input2})
show very clearly the tension between 
the values for $\sin 2\beta$ and $R_b$ in (\ref{input}) and (\ref{input2}),
respectively. We will return to this issue in Section 6. 
For completeness we recall here the even stronger tension that exists between
the value of $R_b$ in (\ref{input2}) and the measured $(\sin
2\beta)_{\phi K_S}=0.47 \pm 0.19$ \cite{BBpage}
coming from the CP asymmetry in $B^0_d(\bar B^0_d)\to \phi K_S$, which is sensitive to new physics in the decay amplitude. 

In  the right panel of Fig.~\ref{fig:Rb_and_gamma} we show $\gamma$ as a function of $\xi$ with $\Delta M_s$ and $\sin 2\beta$ varied in 
the ranges \eqref{CDF} and 
(\ref{input}), respectively.
As the uncertainty in this plot originates dominantly from $\Delta M_s$, 
the main impact of the recent measurement of $\Delta M_s$ in (\ref{CDF})
 is to constrain 
the angle $\gamma$ in the UUT. With the sizable errors on $\xi$ in (\ref{xi})
 and 
$\gamma_\text{true}$ in (\ref{input2}),  the second 
CMFV relation in (\ref{VUBG}) is satisfied, as seen from \eqref{RCMFV} and \eqref{input2}, but clearly this test is not
conclusive at present. It will be interesting to monitor the plots in Fig.~\ref{fig:Rb_and_gamma},  
 when the errors 
on the  values of the quantities involved in these tests will be reduced with
time.

\begin{figure}
\center{\epsfig{file=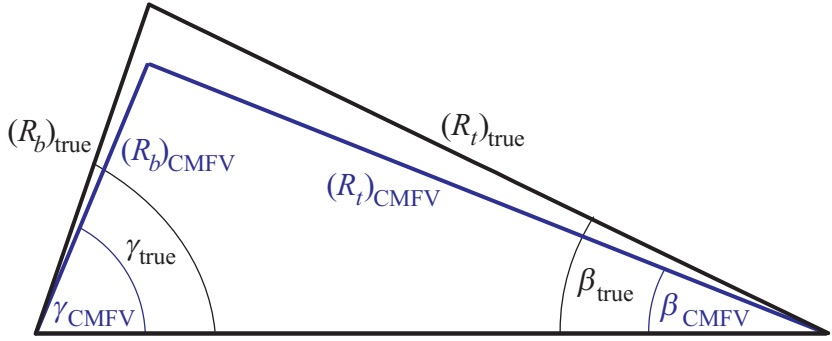}}
\caption{Reference Unitarity Triangle and Universal Unitarity
  Triangle.}
\label{fig:RUT-UUT}
\end{figure}

Finally, in Fig.~\ref{fig:RUT-UUT} we show the universal unitarity triangle
and the reference unitarity triangle, constructed using the central
values in \eqref{input} and \eqref{input2}, respectively. The
qualitative differences between CMFV and tree determination, to which
we will return in Section 6, can
clearly be seen in this figure. However, these differences are small
and the basic message of Fig.~\ref{fig:RUT-UUT} is that from the point
of view of the so-called ``$B_d$-triangle'' of
Fig.~\ref{fig:utriangle}, the present measurements exhibit CMFV in a
reasonable shape.

\section{Implications for Rare Decays}\setcounter{equation}{0}

The result for $\Delta M_s$ in (\ref{CDF}) 
        has immediately 
        four additional profound consequences for CMFV models:
\begin{itemize}
\item
 The ratio
\be\label{R1A}
\frac{Br(B_{s}\to\mu^+\mu^-)}{Br(B_{d}\to\mu^+\mu^-)}
=\frac{\hat B_{B_{d}}}{\hat B_{B_{s}}}
\frac{\tau( B_{s})}{\tau( B_{d})} 
\frac{\Delta M_{s}}{\Delta M_{d}}= 32.4\pm1.9
\ee
can be predicted very accurately \cite{AJB03}, 
subject to only small non-perturbative 
uncertainties in $\hat B_{B_s}/\hat B_{B_d}$ and experimental 
uncertainties in $\tau( B_{s})/\tau( B_{d})$.
\item
Similarly, one can predict
\be\label{Bnunu}
\frac{Br(B\to X_s\nu\bar\nu)}{Br(B\to X_d\nu\bar\nu)}
=\frac{\vts^2}{\vtd^2}
=\frac{m_{B_d}}{m_{B_s}}
\frac{1}{\xi^2}
\frac{\Delta M_s}{\Delta M_d}=22.3\pm2.2,
\end{equation}
where the second relation will offer a very good test of CMFV, once $\vts$ and 
$\vtd$ will be known from the determination of the reference unitarity 
triangle and the error on $\xi$ will be decreased.
\item
From (\ref{Bnunu}) we can also extract
\be\label{vtdvts}
\frac{\vtd}{\vts}=0.212\pm0.011
\ee
which, although a bit larger, is still consistent with the results of the UTfit \cite{UTFIT} and CKMfitter \cite{CKMFIT} 
collaborations and the recent 
determination of this ratio from $B\to V\gamma$ decays \cite{Ball}:
\begin{gather}\label{vtds}
\frac{\vtd}{\vts}_\text{UTfit}=0.202\pm0.008,\qquad 
\frac{\vtd}{\vts}_\text{CKMfitter}=0.2011^{+0.0081}_{-0.0065},\\ 
\frac{\vtd}{\vts}^\text{Belle}_{B\to V\gamma}=0.207\pm0.027(\text{exp.})\pm0.016(\text{th.}),
\end{gather}
where the values given in \eqref{vtds} shifted from $0.198\pm0.010$
and $0.195\pm0.010$, respectively, due to the inclusion of the recent
measurement of $\Delta M_s$ \eqref{CDF} in the analyses.
\item
The branching ratios for $B_{s,d}\to\mu^+\mu^-$ can be predicted within the 
SM and any CMFV model with much higher accuracy than it is 
possible without $\Delta M_{s,d}$. In the SM one has \cite{AJB03}
\be\label{R2}
Br(B_{q}\to\mu^+\mu^-)
=C\frac{\tau(B_{q})}{\hat B_{B_{q}}}
\frac{Y^2(x_t)}{S(x_t)} 
\Delta M_{q}, \qquad (q=s,d)
\ee
with 
\be
C={6\pi}\frac{\eta_Y^2}{\eta_B}
\left(\frac{\alpha}{4\pi\sin^2\theta_{W}}\right)^2\frac{m_\mu^2}{\mw^2}
=4.39\cdot 10^{-10}
\ee
and $S(x_t)=2.33\pm 0.07$ and $Y(x_t)=0.95\pm0.03$ being the relevant top mass dependent
one-loop functions.
\end{itemize}

\begin{figure}
\begin{minipage}{8cm}
\epsfig{file=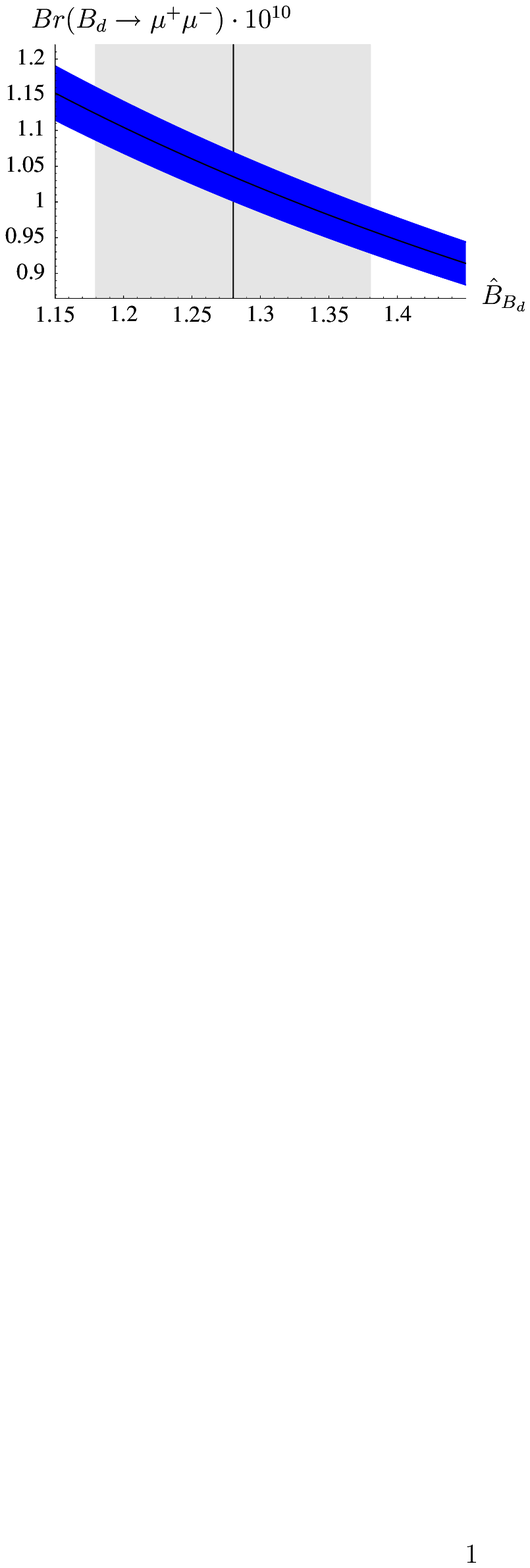,scale=0.9}
\end{minipage}
\hfill
\begin{minipage}{9cm}
\epsfig{file=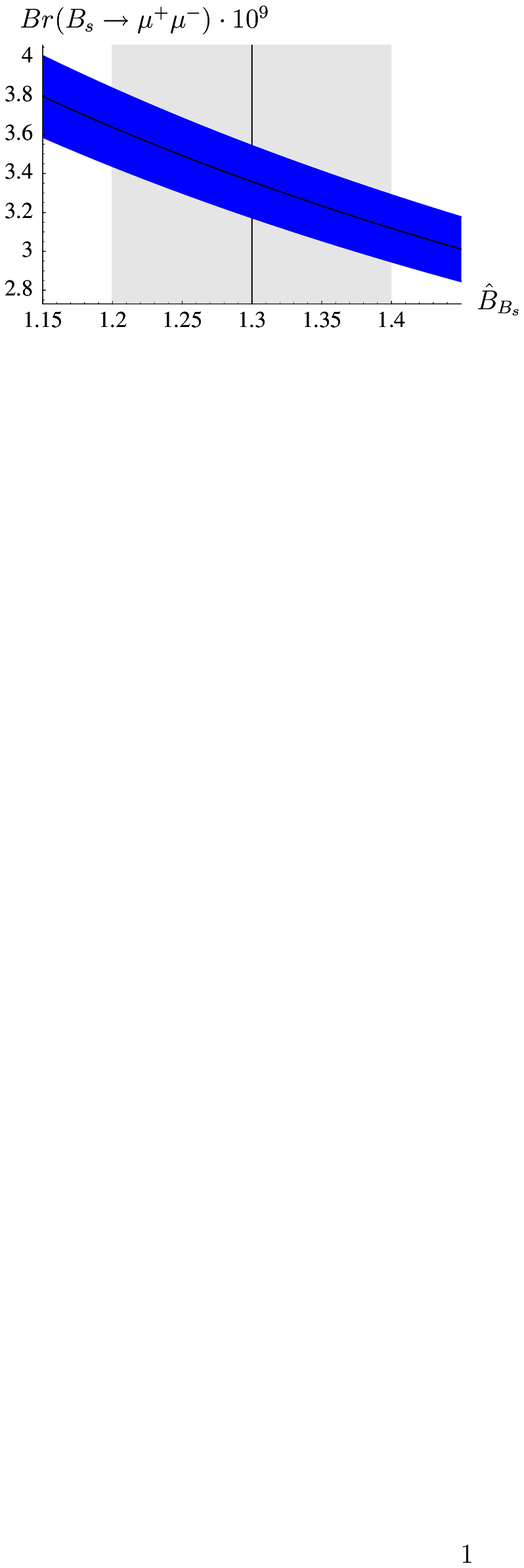,scale=0.9}
\end{minipage}
\caption{$Br(B_d\to\mu^+\mu^-)$ and $Br(B_s\to\mu^+\mu^-)$ in the SM
  as functions of $\hat B_{B_d}$ and $\hat B_{B_s}$, 
respectively.}
\label{fig:SMpred}
\end{figure}

In Fig.~\ref{fig:SMpred} we plot $Br(B_d\to\mu^+\mu^-)$  and 
$Br(B_s\to\mu^+\mu^-)$ in the SM 
as  functions of $\hat B_{B_d}$ and $\hat B_{B_s}$, respectively,  with the
errors in the other quantities entering 
(\ref{R2}) added in quadrature. Clearly, a reduction of the uncertainties on $\hat B_{B_q}$ 
is very desirable.
For $Br(B_d\to\mu^+ \mu^-)$ the updated value  obtained by 
 means of (\ref{R2}) reads
\be\label{SMd}
Br(B_d\to\mu^+ \mu^-)^\text{SM}= (1.03\pm0.09)\cdot 10^{-10},
\ee
and with the value for $\Delta M_s$ in (\ref{CDF}), we also obtain
\be\label{SMs}
Br(B_s\to\mu^+ \mu^-)^\text{SM}= (3.35\pm 0.32)\cdot 10^{-9}.
\ee
These values should be compared with the most  
recent upper bounds 
from CDF \cite{CDF}
\be
Br(B_d\to\mu^+ \mu^-)< 3\cdot 10^{-8}, \qquad
Br(B_s\to\mu^+ \mu^-)< 1\cdot 10^{-7} \quad\quad (95\%~ {\rm C.L.}), 
\ee
implying that there is still a lot of room for new physics 
contributions.

We stress that once LHC is turned on, the accuracy on $\sin 2\beta$ 
and $\Delta M_s$  will match the one of $\Delta M_d$, and consequently
the accuracy of the predicted values for $R_b$ and $\gamma$ 
 in Fig.~\ref{fig:Rb_and_gamma}, 
of the ratios in (\ref{R1A})-(\ref{vtdvts})
and of the SM predictions in (\ref{SMd}) and (\ref{SMs})  
will depend entirely on 
the accuracy of $\xi$ and $\hat B_{B_q}$ which therefore has to be improved.
The resulting numbers from \eqref{R1A}-\eqref{vtdvts} can be considered 
as ``magic numbers of CMFV'' and any deviation of future data 
from these numbers  
will signal new effects beyond CMFV. We underline
the model independent character of these tests. 

Another very important test of CMFV and of MFV in general, still
within $B_{s,d}$ decays, will be the measurement of the mixing induced
asymmetry in $B^0_s(\bar B^0_s)\to\psi\phi$ that is predicted within
the MFV scenario to be $S_{\psi\phi}=0.038\pm0.002$ \cite{UTFIT,CKMFIT}.
We will return to this issue in Section 7.

\section{Tests Beyond $\bm{B_{d,s}}$ Decays}\setcounter{equation}{0}

The tests of CMFV considered so far involve only $B_d$ and $B_s$ 
mesons. Equally important are the tests of the CMFV hypothesis in $K$ meson
decays and even more relevant
those involving correlations between $B$ and $K$ decays that
are implied by CMFV \cite{UUT}.

The cleanest model independent test of MFV in $K$ decays is offered by
$K\to \pi\nu\bar\nu$ decays, 
where the measurement of  $Br(\klpn)$ 
and $Br(\kpn)$ allows a very clean determination of $\sin 2\beta$ 
\cite{BF01,BB4} to be 
compared with the one from $B_d(\bar B_d)\to \psi K_S$. 
The recent NNLO calculation 
of $\kpn$ \cite{BGHN} and the improved calculation of long distance
contributions to this decay \cite{Isidori:2005xm} increased significantly the precision of this test. 
As the  determination of $\sin{2\beta}$ from  $B_d(\bar B_d)\to \psi K_S$
measures the CP-violating phase in $B_d^0-\bar B_d^0$ mixing, while the one
through $K\to\pi\nu\bar\nu$ measures the corresponding phase in $Z^0$-penguin diagrams, it is a very non-trivial MFV test. 
In fact, similarly to $S_{\psi\phi}$, it is a test of the MFV hypothesis and not only of the CMFV one, 
as due to neutrinos in the final state MFV=CMFV in this case.
Unfortunately, due
to slow progress in measuring these two branching ratios, such a test will
only be possible in the next decade.

Thus, for the time being, the only measured quantity in $K$ decays that 
could be used in principle for our purposes is the CP-violating parameter 
$\varepsilon_K$.
As it is the only quantity that is available in the $K^0-\bar K^0$ system, 
its explicit dependence on possible new physics contributions entering 
through the one-loop function $S$ cannot be eliminated within the $K$ 
system alone. For this reason the usual analysis of the UUT involved 
so far only $|V_{ub}/V_{cb}|$, $S_{\psi K_S}$ and the upper bound 
on $\Delta M_d/\Delta M_s$ \cite{UTFIT,BUPAST}.

Here, we would like to point out that in fact the combination of 
$\varepsilon_K$ and $\Delta M_d$, used already in \cite{AJBRB} to derive
a lower bound on $\sin 2\beta$ from CMFV, can also be used in the 
construction of the UUT and generally in the tests of CMFV. Indeed, in all 
CMFV models considered, only the term in $\varepsilon_K$ involving 
$(V_{ts}^*V_{td})^2$ is affected visibly by new physics with the remaining terms 
described by the SM. Eliminating then the one-loop function $S$ in 
$\varepsilon_K$ in terms of $\Delta M_d$ one finds following \cite{AJBRB}
\be\label{Robert}
\sin 2\beta=\frac{0.542}{\kappa}
\left[\frac{|\varepsilon_K|}{\vcb^2\hat B_K}
-4.97\bar\eta P_c(\varepsilon_K)\right]
\ee
with
\be
\kappa=\left[\frac{\Delta M_d}{0.507/\text{ps}}\right]
\left[\frac{214\mev}{F_{B_d}\sqrt{\hat B_{B_d}}}\right]^2, 
\qquad
P_c(\varepsilon_K)=0.29\pm 0.06,
\ee
that should be compared with $\sin 2\beta$ in (\ref{input}). As 
the second term in (\ref{Robert}) is roughly by a factor of three smaller
than the first term, the small model dependence in $\bar\eta$ can be
neglected for practical purposes. The non-perturbative uncertainties in 
$\hat B_K$  and $F_{B_d}\sqrt{\hat B_{B_d}}$ \cite{Hashimoto} 
do not allow a precise test
at present, but the situation could improve in the future.

In summary, 
CMFV has survived its first model independent tests, although there is
some tension between the values of $\beta_\text{true}$ and
$\beta_\text{CMFV}$, as seen in Fig.~\ref{fig:RUT-UUT}. We will return
to this issue in Section 6. Due to the
significant experimental error in  the tree level 
determinations of $\gamma$ and $|V_{ub}/V_{cb}|$ and the theoretical error in 
$\xi$, these tests are not conclusive at present. We are looking forward to the reduction  of these 
 errors. This will allow much more stringent tests of CMFV, 
in particular, 
if in addition 
also the tests of model independent CMFV relations discussed above and in 
\cite{UUT,AJB03,REL} that involve rare $B$ and $K$ decays will also be 
available. 
Future violations of some of these relations would be exciting. 
Therefore, let us ask next what would be the impact of new operators
within MFV on some of the relations discussed above.

\section{The Impact of New Operators}\setcounter{equation}{0}

\begin{figure}
\center{\epsfig{file=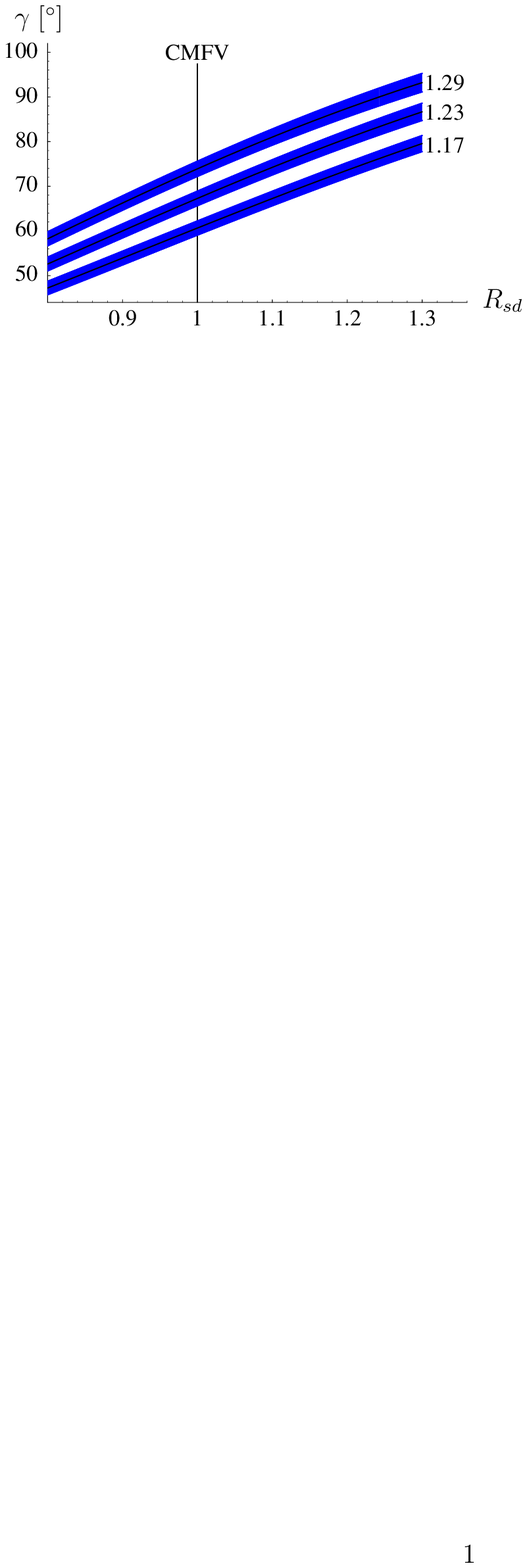}}
\vspace{-0.2cm}
\caption{$\gamma$ as a function of $R_{sd}$ for different values of
  $\xi$.}
\label{fig:Rsd}
\end{figure}

In the most general MFV no new phases beyond the CKM one are allowed
and consequently (\ref{R1}) remains valid. On the other hand in models with 
two Higgs doublets, like the MSSM, new scalar operators originating dominantly 
in Higgs penguin diagrams become important at large $\tan\beta$ and, being 
sensitive to the external masses, modify $\Delta M_d$ and $\Delta M_s$ 
differently \cite{BCRS}
\be\label{DMLT}
\Delta M_q=(\Delta M_q)^{\rm SM} (1+f_q),\quad 
f_{q}\propto -m_b m_{q}\tan^2\beta \qquad (q=d,s).
\ee
Consequently the CMFV relation between $R_t$ and $\Delta M_d/\Delta
M_s$ (\ref{RRt}) is modified to
\be\label{R1C}
R_t=0.923~\left[\frac{\xi}{1.23}\right] 
\sqrt{\frac{17.4/\text{ps}}{\Delta M_s}} 
\sqrt{\frac{\Delta M_d}{0.507/\text{ps}}} \sqrt{R_{sd}},
\qquad R_{sd}=\frac{1+f_s}{1+f_d}.
\ee
In the MSSM at large $\tan\beta$,  
$f_s<0$ and $f_d\approx 0$ \cite{BCRS}, as indicated in (\ref{DMLT}), but as analyzed in 
\cite{AMGIISST}, more generally $f_s$ could also be positive.
In Fig.~\ref{fig:Rsd} we show the impact of $R_{sd}\ne1$ on the value
of $\gamma$ for different values of $\xi$ with the errors in the
remaining quantities added in quadrature. This figure makes clear that
in order to be able to determine  $R_{sd}$ from the data in this
manner, the error in $\xi$ should be significantly reduced.
 
\vspace{0.2truecm}

The new relation in \eqref{R1C} has to be interpreted with some
care. After all, $R_t$ depends only on $\Delta M_d$ and $f_d$ and not on $f_s$ and
$\Delta M_s$, which has been primarily used in (\ref{RRt}) and here to reduce the
non-perturbative uncertainties due to $\hat B_{B_d} F_{B_d}^2$ in $\Delta
M_d$.
For instance, if $f_s$ is indeed negative as found in the MSSM with MFV at
large $\tan \beta$, the measured value of $\Delta M_s$ will also be smaller
cancelling the effect of a negative $f_s$ in calculating $R_t$.
Thus in the MSSM at large $\tan \beta$ in which $f_d \simeq 0$, the numerical
value of $R_t$ is basically not modified with respect to the SM even if
$\Delta M_s$ measured by CDF appears smaller than $(\Delta M_s)^\text{SM}$ as
seen in \eqref{DMsSM}.

The fact that $\Delta M_s$ could indeed be smaller than $(\Delta
M_s)^\text{SM}$ is very interesting, as most MFV models studied in the
literature, with a notable exception of the MSSM at large $\tan \beta$
\cite{BCRS}, predicted $\Delta M_s > (\Delta M_s)^\text{SM}$.
Unfortunately, finding out whether the experimental value of $\Delta
M_s$ is smaller or larger or equal to $(\Delta M_s)^\text{SM}$ would require
a considerable reduction of the uncertainty on $F_{B_s} \sqrt{\hat B_{B_s}}$
that is, at present, roughly $10-15$\%. We will return to this issue in
Section 7.

In this context let us remark that an improved calculation of $F_{B_s}
\sqrt{\hat B_{B_s}}$ together with a rather accurate value of $|V_{ts}|$ and
$\Delta M_s$ would allow to measure in a model independent manner
the function $S(x_t)$ in (\ref{DMsSMb}) and, consequently, to check whether the SM
value of this function agrees with the experimental one.

Of considerable interest is the correlation between new operator effects in 
$\Delta M_s$ and $Br(B_{s,d}\to\mu^+\mu^-)$ that has been pointed out 
in the MSSM with MFV and large $\tan\beta$ in \cite{BCRS} and subsequently 
generalized to arbitrary 
MFV models in \cite{AMGIISST}. In particular within the MSSM, the huge
enhancement of $Br(B_{s,d}\to\mu^+\mu^-)$ at large $\tan\beta$ 
analyzed by many authors in the past
\cite{LTANB} is correlated with the suppression of $\Delta M_s$ with 
respect to the SM, in contrast to the CMFV relation (\ref{R2}). 
Detailed analyses of this correlation can be found in \cite{BCRS,Foster} with 
the most recent ones in \cite{Foster:2006ze,Isidori:2006pk,Carlos}. Here we just want
to remark that due to the 
fact that 
$\Delta M_s$ is found  close to the SM prediction, no large enhancements of 
$Br(B_{d,s}\to \mu^+\mu^-)$ are expected 
within the MSSM with MFV and an observation of $Br(B_{s}\to \mu^+\mu^-)$ and 
$Br(B_{d}\to \mu^+\mu^-)$   with  rates  $\text{few}\cdot10^{-8}$ and 
$\text{few}\cdot10^{-9}$, respectively, would clearly signal new effects beyond 
the MFV framework \cite{MFVB}. Indeed such a correlation between
$\Delta M_s$ and $Br(B_{s}\to \mu^+\mu^-)$  can be avoided in the MSSM with new
sources of flavour violation \cite{Chankowski:2003xd}.

On the other hand, the fact that $\Delta M_s$ has been found below its SM
expectation keeps the MSSM with MFV and large $\tan \beta$ alive and
this version of MSSM would even be favoured if one could convincingly
demonstrate that $\Delta M_s < (\Delta M_s)^\text{SM}$.

Let us remark that in the case of the dominance of scalar operator 
contributions to $Br(B_{d,s}\to \mu^+\mu^-)$, the golden relation (\ref{R1A})
is modified in the MSSM to \cite{AJB03}
\be\label{R1B}
\frac{Br(B_{s}\to\mu^+\mu^-)}{Br(B_{d}\to\mu^+\mu^-)}
=\frac{\hat B_{d}}{\hat B_{s}}
\frac{\tau( B_{s})}{\tau( B_{d})} 
\frac{\Delta M_{s}}{\Delta M_{d}}\left[\frac{m_{B_s}}{m_{B_d}}\right]^4
\frac{1}{1+f_s}
\ee
with $f_s$ being a complicated function of supersymmetric parameters. 
In view of the theoretical cleanness of this relation the measurement of 
 the difference 
between (\ref{R1A}) and (\ref{R1B}) is not out of question. On the 
other hand, the impact of new operators on relation (\ref{Robert}) 
will be difficult to see, as these contributions are small in $\varepsilon_K$ 
and $\Delta M_d$ and the non-perturbative uncertainties involved are still 
significant.

\section{A Brief Look Beyond MFV}\setcounter{equation}{0}

Finally, let us briefly go beyond MFV and admit new flavour
violating interactions, in particular new CP-violating phases as well as
$f_s\not=f_d$. Extensive model independent numerical studies of the UT 
in such general scenarios have been already performed for some time, in 
particular in \cite{UTFIT,CKMFIT,Botella:2005fc,NMFV,Velasco-Sevilla,LLNP,Ligeti:2006pm,Ball:2006xx,Khalil:2006pv,Grossman:2006ce,Datta:2006ne}, 
where references to earlier literature can be found. The analysis of
\cite{NMFV} has recently been updated in \cite{Ligeti:2006pm} in view
of the result in \eqref{CDF}. Here we want to look instead at 
these scenarios in the spirit of the rest of our paper.

Let us then first assume as indicated by the plot in 
Fig.~\ref{fig:Rb_and_gamma} that indeed 
the value of $R_b$  following from (\ref{VUBG}) is smaller  
than the one following from tree level decays. 
While in the case of the angle $\gamma$, nothing conclusive can be 
said at present, let us assume that  $\gamma$ found from tree level 
decays is in the ball park of $75^\circ$, say $\gamma= (75\pm 5)^\circ$, 
that is larger than roughly $60^\circ$ found from the UT fits 
\cite{UTFIT,CKMFIT}.
In fact such large values of $\gamma$ from tree level decays have been 
indicated by the analyses of $B\to \pi\pi$ and $B\to \pi K$ data in 
\cite{SCET,Buras:2005cv}.

In order to see the implications of such findings in a transparent manner, 
let us invert (\ref{VUBG}) to find
\be
R_t=\sqrt{1+R_b^2-2 R_b\cos\gamma},\qquad
\cot\beta=\frac{1-R_b\cos\gamma}{R_b\sin\gamma}.
\ee
In the spirit of the analysis in \cite{Buras:2005cv} we then set 
$\gamma_\text{true}= (75\pm 5)^\circ$ and $(R_b)_\text{true}=0.44\pm0.04$ and determine
the true values of $\beta$ and $R_t$, 
\be\label{true}
\beta_{\rm true}= (25.6\pm 2.3)^\circ ,\qquad   (R_t)_{\rm true}=
0.983\pm 0.038 ,
\ee
to be compared with 
\be\label{CMFV}
\beta_{\rm CMFV}= (21.7\pm1.3)^\circ ,\qquad   (R_t)_{\rm CMFV}=
0.923\pm0.044,
\ee
that follow from (\ref{R1}) and (\ref{RRt}), respectively. The
difference between \eqref{true} and  \eqref{CMFV} is similar to the
one shown in Fig.~\ref{fig:RUT-UUT}, though we have chosen here
$\gamma_\text{true}$ to be larger than the central value in \eqref{input2}.
The present data and the assumption about the true value of $\gamma$ made 
above then imply that
\cite{Buras:2005cv}
\be\label{BMFV}
\beta_{\psi K_S}=\beta_{\rm CMFV} < \beta_{\rm true}, \qquad 
\sin 2(\beta_{\rm true}+\varphi_{B_d}) 
= S_{\psi K_S}, \qquad \varphi_{B_d} < 0
\ee
 with $\varphi_{B_d}$ being a new complex phase, and 
\be\label{BMFV2}
(R_t)_{\rm CMFV} < (R_t)_{\rm true}. 
\ee
The result in \eqref{BMFV} has been first found in
\cite{UTFIT} but the values of $R_t$ and $\gamma$ obtained in  \cite{UTFIT} 
are significantly lower than in \cite{Buras:2005cv} and here. 
The pattern in \eqref{BMFV2} has also been indicated by the analysis 
in \cite{Botella:2005fc}, but we underline that the possible ``discrepancy" in the values 
of $\beta$ is certainly better visible than in the case of $R_t$. 

In particular we find $\varphi_{B_d}= -(3.9\pm2.6)^\circ$ in
agreement with \cite{UTFIT} and  \cite{Buras:2005cv}.
Note that now $\sin 2\beta_{\rm true}= 0.780\pm 0.051$ in conflict with 
$S_{\psi K_S}=0.687\pm0.032$.

The possibility of a new weak phase in $B_d^0-\bar B_d^0$ mixing, 
indicated by (\ref{BMFV}), 
could be tested in other decays sensitive to this mixing but could 
more generally also imply new weak phases in other processes.
The 
latter could then be tested through enhanced CP asymmetries, $S_{\psi\phi}$, 
$A_{\rm CP}(B\to X_s\gamma)$ and $A^{s,d}_{\rm SL}$ that are strongly suppressed in 
MFV models. Such effects could also be clearly seen in $\klpn$.

The origin of a possible  disagreement between $(R_t)_{\rm true}$ and $(R_t)_\text{CMFV}$ is harder to identify 
as it could follow from new flavour violating interactions with the
 same operator structure as in the SM or/and could imply  new enhanced 
operators that are still admitted within the general formulation of 
MFV \cite{AMGIISST} as discussed above. 
Within the $\Delta F=2$ processes alone, it will 
be difficult, if not impossible, to identify which type of violation of 
CMFV  
takes place, unless one specifies a concrete model. On the other hand 
including $\Delta F=1$ transitions in the analysis would allow to 
identify better the origin of the violation of CMFV and MFV relations, 
but such an analysis is clearly  beyond the scope and the spirit
 of our paper.

\section{\boldmath Some Aspects of $S_{ \psi\phi}$ and $A^s_\text{SL}$ \unboldmath}\setcounter{equation}{0}
\label{sec:seven}
In the next years important tests of MFV will come from improved measurements
of the time-dependent mixing induced CP asymmetry
\be
A^s_\text{CP}(\psi\phi,t)= \dfrac{\Gamma(\bar B^0_s(t) \rightarrow 
\psi\phi) - \Gamma(B^0_s(t) \rightarrow  \psi\phi)}{\Gamma(\bar B^0_s(t) \rightarrow \psi\phi) + \Gamma(B^0_s(t) \rightarrow \psi\phi)} \equiv S_{ \psi\phi}
\sin(\Delta M_s t),
\label{7.1}
\ee
where the CP violation in the decay amplitude is set to zero, and of the
semileptonic asymmetry
\be
A^s_\text{SL} = \dfrac{\Gamma(\bar B_s^0 \rightarrow l^+ X) - \Gamma(B_s^0
  \rightarrow l^- X)}{\Gamma(\bar B_s^0 \rightarrow l^+ X) + \Gamma(B_s^0
  \rightarrow l^- X)} =
\text{Im}\left(\dfrac{\Gamma_{12}^s}{M_{12}^s}\right),
\ee
where $\Gamma_{12}^s$ represents the absorptive part of the $B_s^0 - \bar
B_s^0$ amplitude.
The semileptonic asymmetry $A^s_\text{SL}$ has not been measured yet, while
its theoretical prediction in the SM has recently improved thanks to advances
in lattice studies of $\Delta B = 2$ four-fermion operators
\cite{Becirevic:2001xt} and to the NLO perturbative calculations of the corresponding Wilson
coefficients \cite{noi,Beneke:2003az}.

Both asymmetries are very small in MFV models but can be enhanced even by an
order of magnitude if new complex phases are present.
This topic has been extensively discussed in the recent literature, in
particular in  \cite{Ligeti:2006pm} where the
correlation between $A_\text{SL}^s$
and $S_{ \psi\phi}$ has been derived and discussed for the first time.
Here we would like to point out that in most recent papers the sign of the new physics
contribution to $S_{ \psi\phi}$ is incorrect with an evident consequence on
the correlation in question.

Adopting the popular parametrizations of the new physics contributions
\cite{UTFIT,LLNP,Ligeti:2006pm}
\be
\Delta M_s \equiv (\Delta M_s)^\text{SM} |1+h_s e^{2i \sigma_s}| \equiv
(\Delta M_s)^\text{SM} C_{B_s} ,
\label{eq:CBs}
\ee
with
\be
1+h_s e^{2i \sigma_s} \equiv C_{B_s} e^{2i \varphi_{B_s}},
\ee
we find
\be
S_{ \psi\phi} = - \eta_{ \psi\phi} \sin(2\beta_s + 2
\varphi_{B_s})\,,\qquad V_{ts}=-|V_{ts}|e^{-i \beta_s}
\ee
in the parametrization of \cite{UTFIT,LLNP} and
\be
S_{\psi\phi} = - \eta_{ \psi\phi} \left [ h_s \dfrac{\sin 2
    \sigma_s}{C_{B_s}}+\dfrac{\sin{2 \beta_s}(1+h_s \cos 2 \sigma_s)}{C_{B_s}}\right]
\ee
in the parametrization of \cite{Ligeti:2006pm} and setting $\cos 2 \beta_s =
1$, since $\beta_s \simeq -1^\circ$.
Here $\eta_{\psi\phi}$ is the CP parity of the $ \psi\phi$ final state, for
which we take $\eta_{ \psi\phi}=+1$.
We find then
\be\label{7.7}
S_{\psi\phi}=\sin(2|\beta_s|-2\varphi_{B_s}) \approx -\sin 2
\varphi_{B_s},
\ee
or
\be
S_{\psi\phi}=-\dfrac{h_s \sin 2 \sigma_s}{C_{B_s}}+\sin 2 |\beta_s|
\dfrac{1+h_s \cos 2 \sigma_s}{C_{B_s}} \approx -\dfrac{h_s \sin 2
  \sigma_s}{C_{B_s}}.
\ee

While the sign of $(S_{ \psi\phi})^\text{SM}$, obtained from above for
$\sigma_s=0$, $h_s=0$, $C_{B_s}=1$ and $\varphi_{B_s}=0$, agrees with the
recent literature, it is important to clarify that the asymmetry $S_{\psi\phi}$
measures $\sin(2|\beta_s|-2\varphi_{B_s})$ and \emph{not}
$\sin(2|\beta_s|+2\varphi_{B_s})$ as stated in the literature.
This is probably not important for the model independent analysis of
$S_{\psi\phi}$ alone, but it is crucial to have correct signs
when one works with specific new physics models, where the new phase in
$\Delta B=2$ observables is generally correlated with the phases in
$\Delta B=1$ processes, and if different $\Delta B=2$ observables are
considered simultaneously.

\begin{figure}
\center{\epsfig{file=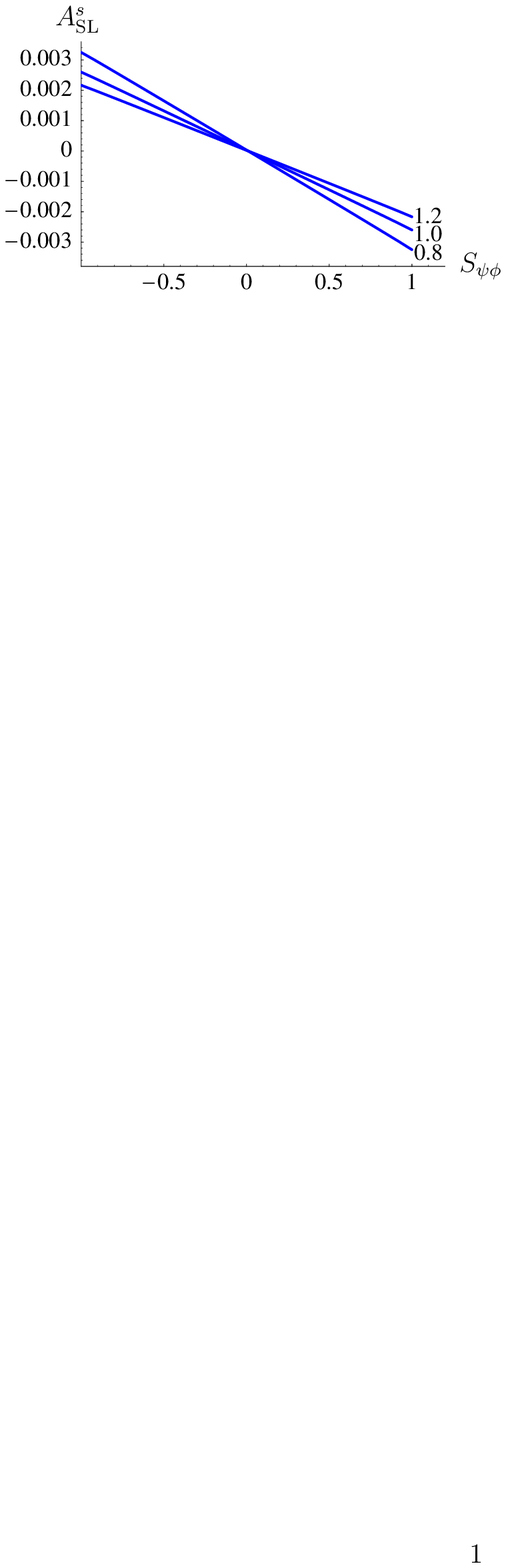}}
\vspace{-0.2cm}
\caption{$A^s_\text{SL}$ as a function of $S_{ \psi\phi}$ for different values
  of $C_{B_s}$.}
\label{fig:ASL}
\end{figure}

As an example let us consider $A^s_\text{SL}$, that can be rewritten as
\bea
A^s_\text{SL} &=& \text{Im}\left(\dfrac{\Gamma^s_{12}}{M^s_{12}}\right)^\text{SM}\dfrac{\cos
  2\varphi_{B_s}}{C_{B_s}} -
\text{Re}\left(\dfrac{\Gamma^s_{12}}{M^s_{12}}\right)^\text{SM}\dfrac{\sin 2\varphi_{B_s}}{C_{B_s}}\nn\\
 &\approx&  - \text{Re}\left(\dfrac{\Gamma^s_{12}}{M^s_{12}}\right)^\text{SM}\dfrac{\sin
  2\varphi_{B_s}}{C_{B_s}}.
\eea
Recalling that $\text{Re}(\Gamma^s_{12}/M^s_{12})^\text{SM} <0$ and using \eqref{7.7},
we find the following correlation between $A^s_\text{SL}$ and $S_{ \psi\phi}$
\be
A^s_\text{SL} = -\left |
 \text{Re}\left(\dfrac{\Gamma^s_{12}}{M^s_{12}}\right)^\text{SM}\right|
\dfrac{1}{C_{B_s}}S_{ \psi\phi},
\label{ASLcorr}
\ee
shown in Fig. \ref{fig:ASL}, for different values of $C_{B_s}$ and with
$|\text{Re}(\Gamma^s_{12}/M^s_{12})^\text{SM}|=(2.6 \pm 1.0)\cdot 10^{-3}$
\cite{noi} fixed to its central value.
We would like to stress that already a rather small value of $S_{\psi \phi}
\simeq 0.1$ would lead to an order of magnitude enhancement of $A^s_\text{SL}$
relative to its SM expectation.

We note that the theoretical prediction for
$\text{Re}(\Gamma^s_{12}/M^s_{12})^\text{SM}$ obtained in \cite{noi} and used
here is smaller than the value found in \cite{Beneke:1998sy}. 
This difference is mainly due to the contribution of
$\mathcal{O}(1/m_b^4)$ in the Heavy Quark Expansion (HQE), which in \cite{Beneke:1998sy} is
wholly estimated in the vacuum saturation approximation (VSA), while in
\cite{noi} the matrix elements of two dimension-seven operators are expressed
in terms of those calculated on the lattice. 
Moreover, we emphasize that the negative sign in (\ref{ASLcorr}), now
confirmed also in \cite{Ligeti:2006pm}, is model independent as
$C_{B_s}=|1 + h_s \exp{(2 i \sigma_s)}|>0$.
In \cite{Ligeti:2006pm} the effect of $C_{B_s}$ is enclosed in
$\mathcal{O}(h_s^2)$ corrections, as an expansion in $h_s$ is performed.

Strictly speaking the formula (\ref{ASLcorr}) is not a correlation between 
 $A^s_\text{SL}$ and $S_{\psi \phi}$ only, but a triple correlation between these two quantities and $C_{B_s}$. It is so general that it cannot be used as a test of any 
 extension of the SM but in any model the knowledge of two among these 
 three quantities allows to predict the third one.
Therefore we would like to point out that (\ref{ASLcorr}) offers
in principle an alternative way to find out whether $\Delta M_s$ differs from $(\Delta M_s)^\text{SM}$.
Indeed, the inversion of (\ref{ASLcorr}) together with (\ref{eq:CBs}) yields
\be
\dfrac{\Delta M_s}{(\Delta M_s)^\text{SM}}= -\left |
 \text{Re}\left(\dfrac{\Gamma^s_{12}}{M^s_{12}}\right)^\text{SM}\right|
\dfrac{S_{ \psi\phi}}{A^s_\text{SL}}\,.
\label{eq:DMsratio}
\ee
With respect to $(\Delta M_s)^\text{SM}$,
$\text{Re}(\Gamma^s_{12}/M^s_{12})^\text{SM}$ is free from the uncertainty coming
from the decay constant $F_{B_s}$.
On the other hand, in $\text{Re}(\Gamma^s_{12}/M^s_{12})^\text{SM}$ significant
cancellations occur at NLO and at $\mathcal{O}(1/m_b^4)$ in the HQE, which make it sensitive to the
dimension-seven operators, whose most matrix elements have never been
estimated out of the VSA.
Future lattice calculations together with experimental measurements of 
 the semileptonic asymmetry $A^s_\text{SL}$ are certainly desired for a
 significant determination of $\Delta M_s/(\Delta M_s)^\text{SM}$ through
 (\ref{eq:DMsratio}).  

Similarly, one has in the $B_d$ system
\be
\dfrac{\Delta M_d}{(\Delta M_d)^\text{SM}}= \left |
 \text{Re}\left(\dfrac{\Gamma^d_{12}}{M^d_{12}}\right)^\text{SM}\right|
\dfrac{\sin 2 \varphi_{B_d}}{A^d_\text{SL}}+\text{Im}\left(\dfrac{\Gamma^d_{12}}{M^d_{12}}\right)^\text{SM}
\dfrac{\cos 2 \varphi_{B_d}}{A^d_\text{SL}}\,,
\label{eq:DMdratio}
\ee
where $\varphi_{B_d}$ is the new phase in (\ref{BMFV}).
We note that in this case $\text{Im}(\Gamma^d_{12}/M^d_{12})^\text{SM}= - (6.4
\pm 1.4)\cdot 10^{-4}$ cannot be neglected with respect to
$|\text{Re}(\Gamma^d_{12}/M^d_{12})^\text{SM}|=(3.0 \pm 1.0)\cdot 10^{-3}$
\cite{noi}.
Finally,  one could use
\be
\frac{\Delta M_q}{(\Delta M_q)^\text{SM}} = 
-\left(\dfrac{\Delta M_q}{\Delta \Gamma_q}\right) \text{Re}\left(\dfrac{\Gamma^q_{12}}{M^q_{12}}\right)^\text{SM} \cos 2\varphi_{B_q}\,,
\label{eq:DMqr}
\ee
with $\varphi_{B_q}$ extracted from $S_{\psi\phi}$ and $S_{\psi K_S}$ 
for $q=s$ and $q=d$, respectively.
These proposals have been recently adopted in \cite{Blanke:2006sb} where an extensive
phenomenological analysis in the Littlest Higgs Model with T-parity has been performed.
It remains to be seen whether in the future our proposals to measure the ratios $\Delta
M_q/(\Delta M_q)^\text{SM}$ by means of (\ref{eq:DMsratio})-(\ref{eq:DMqr})
will be more effective than the direct calculations of $(\Delta M_q)^\text{SM}$.
\section{Conclusions}\setcounter{equation}{0}

The recent measurements of $\Delta M_s$ by the CDF and D{\O}
collaborations gave another support to the hypothesis of MFV. Even if
possible signals of non-MFV interactions, like $\varphi_{B_d}\ne 0$
and $(R_t)_\text{CMFV}<(R_t)_\text{true}$, are indicated by the data,
they are small as seen in Fig.~\ref{fig:RUT-UUT}. However, it should
be emphasized that future measurements of CP violation in $B_s$
decays, in particular of the CP asymmetries $S_{\psi\phi}$ and
$A^s_{\rm SL}$ and of the branching ratios
$Br(B_{d,s}\to\mu^+\mu^-)$, could modify our picture of non-MFV effects
significantly. Also the signals of new weak phases in $B\to\pi K$
decays, discussed in  \cite{Buras:2005cv} and references therein,
should not be forgotten.

In the present paper we have concentrated on quantities like ratios of 
 branching ratios, $\Delta M_d/\Delta M_s$ and various CP asymmetries 
 which do not require the direct use of the weak decay constants 
 $F_{B_q}$ that are plagued by large non-perturbative uncertainties. 
 Observables sensitive only to $\xi$ and $\hat B_{B_q}$ have a better 
 chance to help us in identifying new physics contributions. One of 
 the important tasks for the coming years will be to find out whether 
the data favour positive or negative new physics contributions to 
$\Delta M_q$. As seen in (1.3), from the present perspective, this 
will not be soon possible through a direct calculation of $\Delta M_q$. 
Therefore, we have proposed the formulae (\ref{eq:DMsratio})-(\ref{eq:DMqr}) as alternative 
ways to shed light on this important question. We are aware that also these 
routes are very challenging but they definitely should be followed once the 
data on $A^q_\text{SL}$ and improved data on $\Delta\Gamma_q$ 
will be available.

Truly exciting times are coming for MFV. We should be able to decide
in about $2-3$ years, whether this simple hypothesis survived all
model independent tests summarized in this paper, with the final
precise tests of the correlations between $B$ and $K$ systems left for
$K\to\pi\nu\bar\nu$ in the first years of the next decade. 
On the other hand if non-MFV interactions will be signalled by the 
data, flavour physics will be even more exciting.
We hope
that the formulae and plots collected above will help in monitoring
these events in a transparent manner.
 
\vspace{0.5truecm}

\noindent {\bf Acknowledgements}

\noindent We would like to thank Ulrich Haisch, Luca Silvestrini, Selma Uhlig,
Andreas Weiler and Daria Zieminska for critical comments and illuminating discussions.
This research was partially supported by the German `Bundesministerium f\"ur 
Bildung und Forschung' under contract 05HT4WOA/3. D.G. acknowledges the support of `Fondazione Della Riccia', Firenze, Italy.

\renewcommand{\baselinestretch}{0.95}

\end{document}